\documentclass[a4paper,11pt]{article}
\usepackage{pos}
\usepackage{orcidlink}
\usepackage{soul}
\usepackage[normalem]{ulem}
\usepackage{subcaption}
\usepackage{multirow}

\title{Precision determination of $B_c$ and $B_c^*$
 meson decay constants from lattice QCD}
\ShortTitle{$B_c$ and $B_c^*$
 decay constants from lattice QCD}

\author*[a]{Navdeep Singh Dhindsa~\orcidlink{0000-0002-3133-6979}}

\author[a]{Debsubhra Chakraborty~\orcidlink{0000-0001-5815-4182}}

\author[a]{Nilmani Mathur~\orcidlink{0000-0003-2422-7317}}

\affiliation[a]{Department of Theoretical Physics, Tata Institute of Fundamental Research,\\Homi Bhabha Road, Mumbai 400005, India}

\emailAdd{navdeep@theory.tifr.res.in}
\emailAdd{navdeep.s.dhindsa@gmail.com}

\FullConference{The 43rd International Symposium on Lattice Field Theory (Lattice 2026)\\
July 26 to August 1, 2026\\
University of Maryland, College Park, USA\\}

\abstract{
We present a lattice QCD study of heavy-heavy meson decay constants using the HISQ gauge ensemble generated by the MILC Collaboration. Employing the HISQ action for both charm and bottom valence quarks tuned to their physical masses, we determine pseudoscalar, vector, and tensor decay constants for heavy quarkonium and $B_c$ systems. Preliminary results are presented for the decay constants of the $B_c$, $B_c^*$, $\eta_c$, $\eta_b$, and $\Upsilon$ mesons.
}

\begin{document}
\maketitle

\section{Introduction}
\label{sec:intro}
Precise determinations of the Cabibbo--Kobayashi--Maskawa matrix element $|V_{cb}|$ remain an important challenge in flavour physics. Inclusive and exclusive determinations continue to show a persistent tension \cite{FlavourLatticeAveragingGroupFLAG:2024oxs}, placing increased emphasis on improved control of the hadronic quantities entering semileptonic $B$-meson decays. Complementary information on the underlying heavy-quark dynamics can be obtained from the $B_c$ system, in which the bottom and charm flavours are simultaneously present in a single bound state. As the only meson composed of two different heavy quarks, the $B_c$ represents a unique unequal-mass heavy-heavy bound state involving the two distinct scales $m_b$ and $m_c$. In contrast to charmonium and bottomonium, the $B_c$ carries open flavour quantum numbers and decays weakly, while at the same time retaining many of the characteristic features of a heavy-quark bound state. It therefore provides a particularly interesting laboratory for studying two-scale heavy-quark dynamics.

The purely leptonic decay $B_c\to \ell\nu$ is theoretically clean: all nonperturbative strong-interaction effects are encoded in the single decay constant $f_{B_c}$. Purely leptonic $B_c$ decays remain experimentally challenging. Still, the growing experimental study of the $B_c$ system and the prospect of larger data samples at current and future facilities provide increasing motivation for precise predictions of quantities such as $f_{B_c}$. In addition, $f_{B_c}$ can provide useful input to dispersive and unitarity constraints on $B\to D^{(*)}$ form factors, offering complementary information relevant to exclusive determinations of $|V_{cb}|$.

The vector partner $B_c^*$ is also of particular phenomenological interest. Its vector and tensor decay constants enter dispersive unitarity constraints on $b\to c$ form factors. The tensor decay constant $f_{B_c^*}^{T}$ is especially relevant in searches for physics beyond the Standard Model, since tensor interactions, which are absent in the Standard Model, can arise in extensions of the weak charged-current interaction. A precise determination of $f_{B_c^*}^{T}$ therefore provides useful nonperturbative input for phenomenological studies of possible tensor contributions to $b\to c\ell\nu$ transitions.

A quantitative determination of these decay constants requires nonperturbative control of QCD while simultaneously resolving the charm- and bottom-quark scales. Despite their phenomenological interest, lattice determinations of these quantities remain relatively limited. To date, only a handful of unquenched lattice calculations are available for $f_{B_c}$ \cite{McNeile:2012qf, Colquhoun:2015oha, Becirevic:2018qlo, Cai:2026xja} and $f_{B_c^*}$ \cite{Colquhoun:2015oha, Becirevic:2018qlo, Cai:2026xja}, while no lattice determination of the tensor decay constant $f_{B_c^*}^{T}$ is currently available. The existing results also show a noticeable spread in their central values; for example, determinations of $f_{B_c}$ range from $396(12)$ MeV to $451.0(5.2)$ MeV, with the recent CLQCD calculation \cite{Cai:2026xja} using multiple lattice spacings and reporting a value at the upper end of the existing determinations. This spread provides additional motivation for an independent and accurate determination of these quantities. Here, we use an ultrafine lattice, where both charm and bottom quarks are treated fully relativistically, to obtain high-precision determinations of $f_{B_c}$ and $f_{B_c^*}$, together with the first lattice determination of the tensor decay constant $f_{B_c^*}^{T}$. The numerical setup is described in Sec.~\ref{sec:setup}, followed by the preliminary results in Sec.~\ref{sec:results}, with a summary and outlook in Sec.~\ref{sec:summary}.

\section{Numerical Setup}
\label{sec:setup}

We use a gauge ensemble with $N_f=2+1+1$ flavours of highly improved staggered quarks (HISQ), generated by the MILC Collaboration \cite{Bazavov:2017lyh}. The ensemble has a lattice spacing of $a=0.03271(20)\,\mathrm{fm}$ \cite{Hatton:2020qhk} and a volume of $96^3\times288$. For the valence quarks, we use the standard HISQ action with Fat7-smeared and reunitarized gauge links, including the Lepage correction and the three-link Naik term. The mass-dependent correction to the Naik term is included for the heavy valence quarks, with the correction calculated at tree level. The present analysis uses 400 gauge configurations with one random color-wall source on a gauge-fixed time slice for each configuration. The ensemble parameters and valence-quark masses are summarised in Table~\ref{tab:ensemble}.

\begin{table}[t]
\centering
\small
\begin{tabular}{cccc}
\hline\hline
\multicolumn{4}{c}{Ensemble} \\
\hline
$a$ (fm) & $L^3\times T$ & $M_\pi$ (MeV) 
& $N_{\rm cfg}\times N_{\rm src}$ \\
\hline
$0.03271(20)$ & $96^3\times288$ & $309$ & $400\times1$ \\
\hline
\hline
\end{tabular}

\vspace{3mm}

\begin{tabular}{ccccc}
\hline
\hline
\multicolumn{3}{c}{Sea-quark masses} &
\multicolumn{2}{c}{Valence-quark masses} \\
\hline
$am_l^{\rm sea}$ & $am_s^{\rm sea}$ & $am_c^{\rm sea}$
& $am_c^{\rm val}$ & $am_b^{\rm val}$ \\
\hline
$0.00223$ & $0.01115$ & $0.1316$ & $0.1370$ & $0.6223$ \\
\hline\hline
\end{tabular}
\caption{Parameters of the $N_f=2+1+1$ MILC HISQ ensemble used in this work.}
\label{tab:ensemble}
\end{table}

This is the finest MILC HISQ ensemble currently available, allowing the bottom quark to be treated relativistically with the HISQ action while keeping $am_b<1$. Both charm and bottom quarks can therefore be described within the same relativistic lattice formulation. At this ultrafine lattice spacing, taste-breaking effects are expected to be suppressed, and the use of the HISQ action for both sea and valence quarks avoids mixed-action systematics. The tuning of the valence charm and bottom masses follows the determinations from the $J/\psi$ \cite{Dhindsa:2024erk} and the $\Upsilon$ systems \cite{Radhakrishnan:2026ptj}, respectively. 

To extract the ground-state masses and decay amplitudes, we use local 0-link spin--taste bilinears \cite{Golterman:1985dz, Follana:2006rc}. For the pseudoscalar, vector, and tensor channels, the relevant operators are
\begin{equation}
\mathcal{O}^{P}
 = \bar{\psi}_{b}(\gamma_{5}\otimes\xi_{5})\psi_{c},
\qquad
\mathcal{O}^{V}_{i}
 = \bar{\psi}_{b}(\gamma_{i}\otimes\xi_{i})\psi_{c},
\qquad
\mathcal{O}^{T}_{i}
 = \bar{\psi}_{b}(\gamma_{i4}\otimes\xi_{i4})\psi_{c},
\qquad i=x,y,z .
\label{eq:operators}
\end{equation}
Here, the $\gamma$ and $\xi$ matrices act on the spin and taste degrees of freedom, respectively, with $\gamma_{i4}\equiv\gamma_i\gamma_4$ and $\xi_{i4}\equiv\xi_i\xi_4$. For the vector and tensor channels, all three spatial components are calculated and averaged.

For each channel, the ground-state properties are extracted from the corresponding zero-momentum two-point correlation function,
\begin{equation}
C_{\mathcal O}(t)
=
\sum_{\mathbf{x}}
\left\langle
\mathcal O(\mathbf{x},t)\,
\mathcal O^\dagger(\mathbf{x}_{\rm src},t_{\rm src})
\right\rangle ,
\label{eq:twopoint}
\end{equation}
where $\mathcal O$ denotes the appropriate interpolating operator given in Eq.~\eqref{eq:operators}. The source operator is constructed using a random color-wall source, consisting of three independent random color-vector fields distributed over the source time slice. 

For staggered fermions, the meson correlators couple not only to the desired states but also to their opposite-parity partners. At a finite temporal extent $T$, the corresponding spectral decomposition can be
written as
\begin{equation}
C(t)
=
\sum_n A_n
\left[
e^{-E_n\tau}+e^{-E_n(T-\tau)}
\right]
-
(-1)^\tau
\sum_n B_n
\left[
e^{-E_n^{\,o}\tau}+e^{-E_n^{\,o}(T-\tau)}
\right],
\label{eq:staggered_corr}
\end{equation}
where $\tau=t-t_{\rm src}$, while $E_n$ and $E_n^{\,o}$ denote the energies of the ordinary and opposite-parity states, respectively. The source time is fixed at $t_{\rm src}=0$ throughout, so that
$\tau=t-t_{\rm src}=t$. The factor $(-1)^\tau$ gives rise to the characteristic even--odd
oscillation of the staggered correlators.

To reduce these oscillations before the ground-state analysis, we apply a local three-point geometric smoothing~\cite{Dhindsa:2024erk}, 
\begin{equation}
C_{\rm sm}(t)
=
\sqrt{
C(t)\sqrt{C(t-1)C(t+1)}
}\, .
\label{eq:smoothing}
\end{equation}
The resulting smoothed correlators are then used to extract the ground-state masses and decay amplitudes. As a cross-check of the smoothing procedure, we also perform independent fits to the even and odd time slices of the unsmoothed correlators. The resulting ground-state quantities are consistent with those obtained from the smoothed correlators, indicating that the smoothing does not introduce a bias in the extracted results. The analysis described above forms the basis for the results presented in the next section.

\section{Results}
\label{sec:results}

The smoothed correlators introduced in Eq.~\eqref{eq:smoothing} are analysed in the first half of the temporal extent and fitted over the range $t\in[t_{\min},t_{\max}]$, with $t_{\max}<T/2$, using the single-state form
\begin{equation}
C_{\rm sm}(t)=A e^{-Et},
\label{eq:groundfit}
\end{equation}
from which the ground-state energy $E$ and amplitude $A$ are extracted. The stability of the extracted quantities is examined by varying $t_{\min}$ while keeping $t_{\max}$ fixed.

The pseudoscalar and vector decay constants are conventionally defined through
\begin{align}
\langle 0|\bar b\gamma_\mu\gamma_5 c|B_c(p)\rangle
 &= i f_{B_c}p_\mu ,
\\
\langle 0|\bar b\gamma_\mu c|B_c^*(p,\lambda)\rangle
 &= f_{B_c^*}M_{B_c^*}\epsilon_\mu^{(\lambda)} ,
\end{align}
while the tensor decay constant is defined by
\begin{equation}
\langle 0|\bar b\sigma_{\alpha\beta}c|B_c^*(p,\lambda)\rangle
=
i f_{B_c^*}^{T}(\mu)
\left[
\epsilon_\alpha^{(\lambda)}p_\beta
-
\epsilon_\beta^{(\lambda)}p_\alpha
\right].
\end{equation}

At zero momentum, the fitted ground-state amplitude is related to the
corresponding matrix element through the spectral decomposition
\begin{equation}
A=\frac{|\langle 0|\mathcal{O}|H\rangle|^2}{2E_H},
\end{equation}
where $\mathcal{O}$ denotes the corresponding interpolating operator, $H$ the ground-state hadron, and $E_H=M_H$ at zero momentum. Using this relation together with the axial Ward identity, the pseudoscalar decay constant is obtained directly from the fitted amplitude without an additional current-renormalisation factor,
\begin{equation}
f_{B_c}
=
(am_b+am_c)
\sqrt{\frac{2A_P}{E_P^3}}\,a^{-1}.
\label{eq:fps}
\end{equation}
For the vector and tensor channels, the corresponding quantities prior to current renormalisation are
\begin{equation}
\frac{f_{B_c^*}}{Z_V}
=
\sqrt{\frac{2A_V}{E_V}}\,a^{-1},
\qquad
\frac{f_{B_c^*}^{T}}{Z_T}
=
\sqrt{\frac{2A_T}{E_T}}\,a^{-1},
\label{eq:fvt}
\end{equation}
where $Z_V$ and $Z_T$ denote the vector- and tensor-current renormalisation factors, respectively.

\begin{figure}[h!]
    \centering
    \begin{subfigure}[t]{0.49\textwidth}
        \centering
        \includegraphics[width=\textwidth]{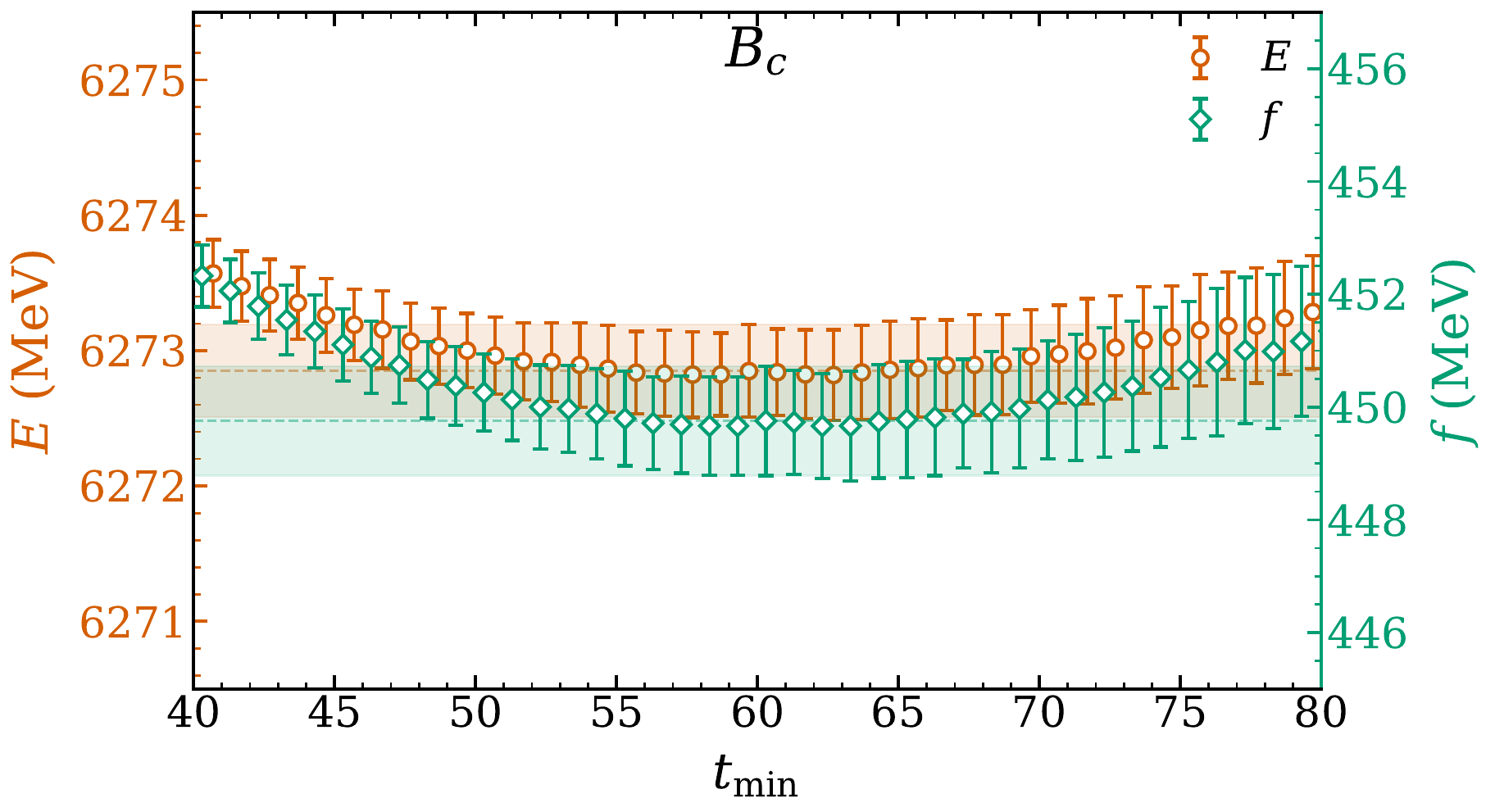}
        \caption{$B_c$ pseudoscalar channel; the chosen fit uses $t_{\min}=60$.}
        \label{fig:tmin_bc}
    \end{subfigure}
    \hfill
    \begin{subfigure}[t]{0.49\textwidth}
        \centering
        \includegraphics[width=\textwidth]{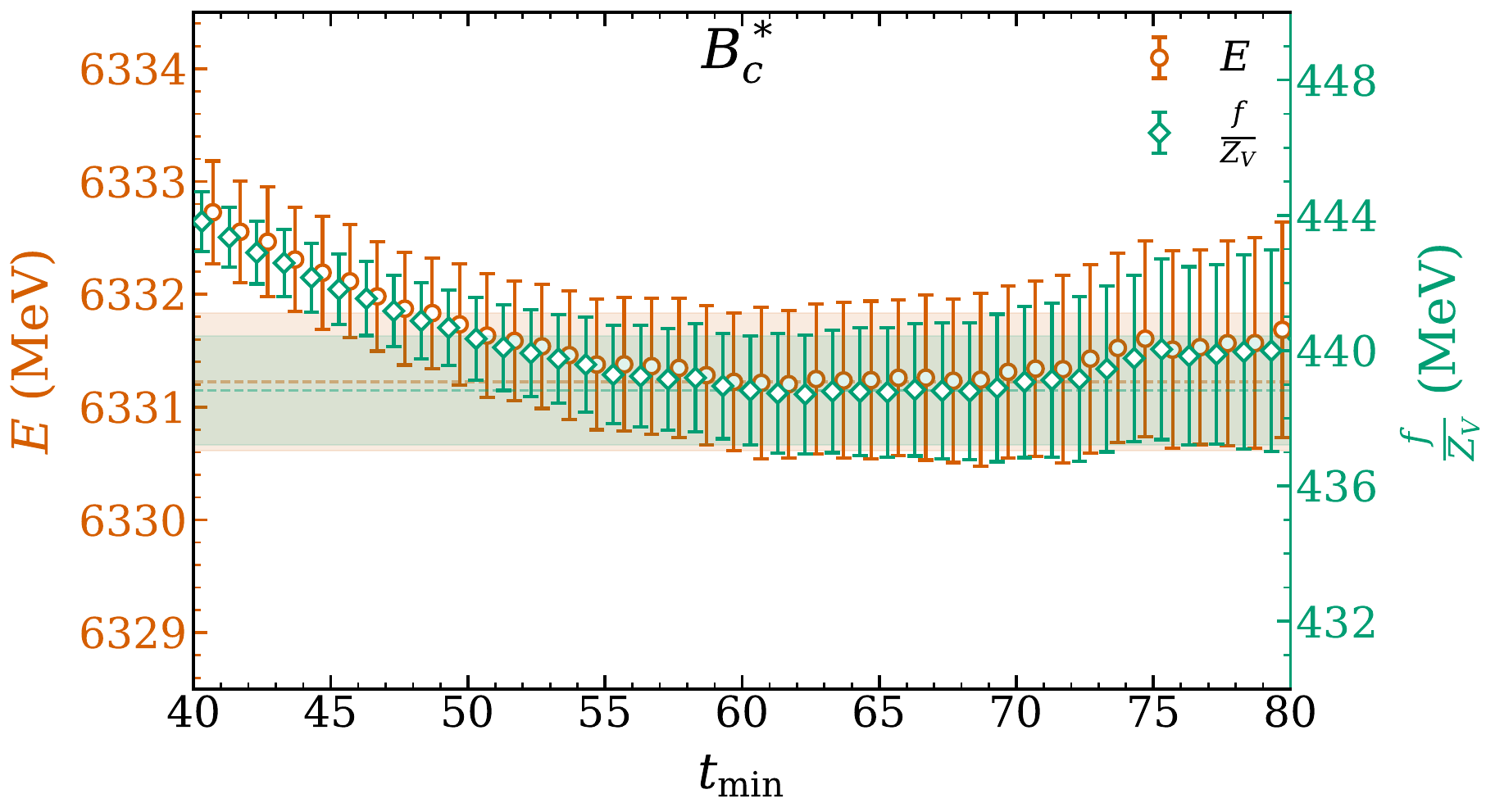}
        \caption{$B_c^*$ vector channel; the chosen fit uses $t_{\min}=60$.}
        \label{fig:tmin_bcstar}
    \end{subfigure}
   
    \begin{subfigure}[t]{0.49\textwidth}
        \centering
        \includegraphics[width=\textwidth]{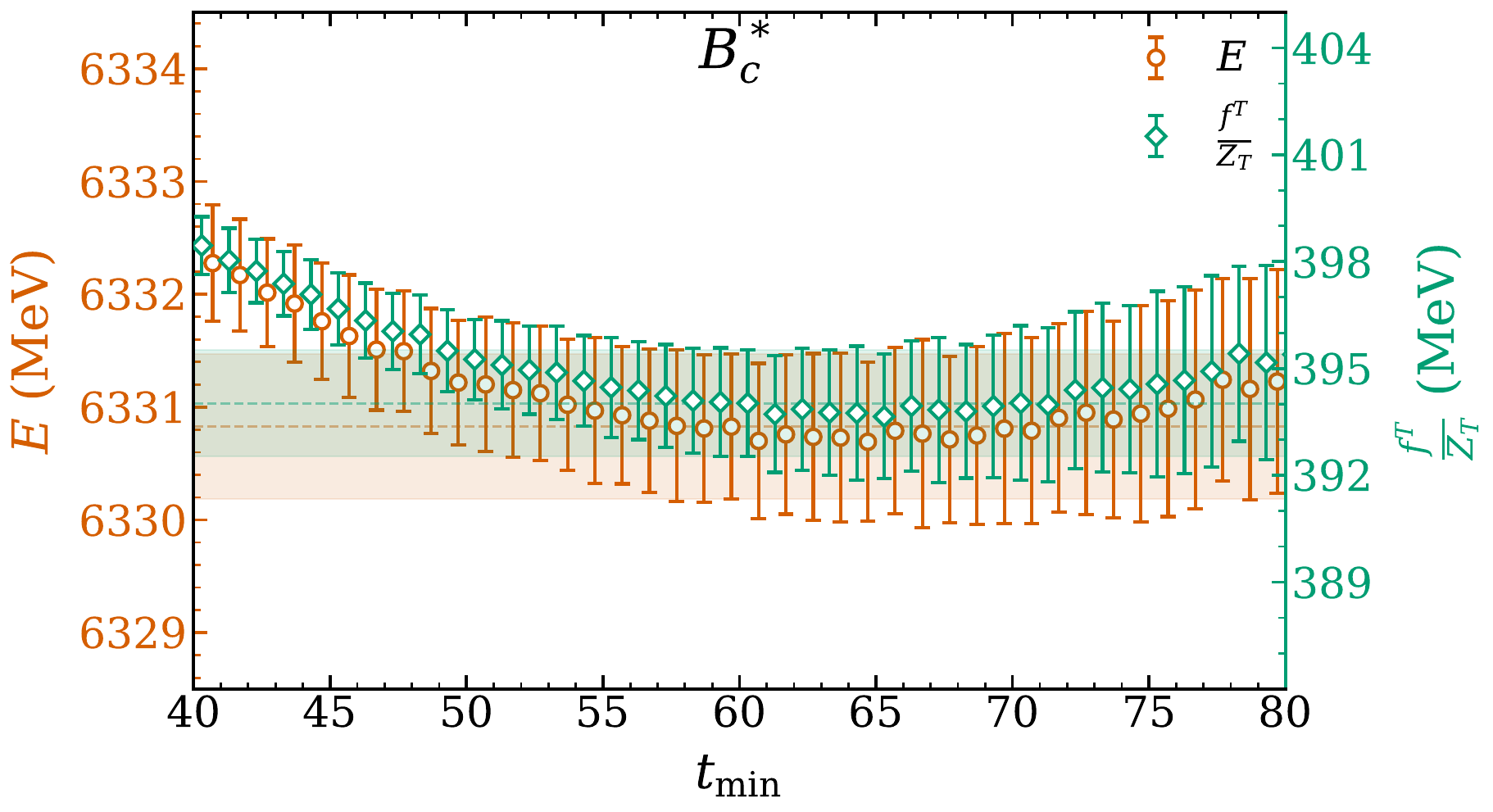}
        \caption{$B_c^*$ tensor channel; the chosen fit uses $t_{\min}=60$.}
        \label{fig:tmin_bcstar_tensor}
    \end{subfigure}

    \caption{$t_{\min}$ dependence of the extracted $B_c$ and $B_c^*$ masses and decay quantities. For the vector and tensor channels, the unrenormalised combinations $f_{B_c^*}/Z_V$ and $f_{B_c^*}^{T}/Z_T$ are shown. The upper end of the fit window is
    fixed to $t_{\max}=120$ throughout.}

    \label{fig:tmin}
\end{figure}

We fit the folded correlators with a single exponential at different fit windows \{$t_{\min}, t_{\max}$\}. Figure~\ref{fig:tmin} shows the dependence of the extracted masses and decay quantities on $t_{\min}$, with $t_{\max}=120$ held fixed. Results are shown for the pseudoscalar $B_c$ channel and for the vector and tensor $B_c^*$ channels over a range of $t_{\min}$ values. The observed $t_{\min}$ dependence is used to assess the dependence of the ground-state extraction to the choice of fit window and to select the best fit ranges with a proper systematics.

In the present proceedings, the vector and tensor results are shown before applying the corresponding current-renormalisation factors. The uncertainties shown are statistical only and do not yet include a complete assessment of systematic uncertainties, such as those associated with the choice of fit window, heavy-quark mass tuning, scale setting, discretisation effects, and other corrections. A more complete analysis, including increased statistics, current renormalisation, a detailed systematic error budget, and final determinations of the decay constants, will be presented in a forthcoming publication. The preliminary values obtained from the adopted fit ranges are summarised in Table~\ref{tab:results}.  Along with the $B_c$ and $B_c^*$ results, we also quote the corresponding quantities for $\eta_c$, $\eta_b$, and the $\Upsilon$ in both the vector and tensor channels. 

\begin{table}[t]
\centering
\small
\begin{tabular}{lccc}
\hline\hline
Channel
& Mass [MeV]
& Decay quantity
& Value [MeV] \\
\hline

$\eta_c$
& $2983.55(30)$
& $f_{\eta_c}$
& $396.41(81)$ \\

$\eta_b$
& $9404.07(14)$
& $f_{\eta_b}$
& $722.76(68)$ \\

$B_c$
& $6272.85(34)$
& $f_{B_c}$
& $449.76(97)$ \\

$B_c^*$ (vector)
& $6331.22(61)$
& $f_{B_c^*}/Z_V$
& $438.82(161)$ \\

$B_c^*$ (tensor)
& $6330.82(64)$
& $f_{B_c^*}^{T}/Z_T$
& $394.02(150)$ \\

$\Upsilon$ (vector)
& $9458.14(18)$
& $f_{\Upsilon}/Z_V$
& $700.30(79)$ \\

$\Upsilon$ (tensor)
& $9458.27(17)$
& $f_{\Upsilon}^{T}/Z_T$
& $646.47(70)$ \\
\hline\hline
\end{tabular}
\caption{Preliminary ground-state masses and decay quantities obtained in this analysis. The uncertainties shown are statistical only. For the vector and tensor channels, the quoted decay quantities are shown before applying the corresponding current-renormalisation factors.}
\label{tab:results}
\end{table}

It is useful to compare these preliminary results with the recent CLQCD calculation \cite{Cai:2026xja}. Our $B_c$ decay constant is in close agreement with their determination, while the unrenormalised $B_c^*$ result is also very similar in magnitude. Since the vector-current renormalisation is expected to provide only a small correction, we anticipate a close correspondence for the renormalised result as well.

It is also useful to compare our preliminary results with the recent QCD sum-rule analysis of Ref.~\cite{CostaiReina:2026deh} and the references therein, which provides determinations of the pseudoscalar and vector decay constants as well as the first QCD sum-rule determination of $f_{B_c^*}^{T}$. Their results for $f_{B_c}$ and $f_{B_c^*}$ are compatible with our preliminary determinations within their quoted uncertainties. For the tensor decay constant, their central value is higher than our present unrenormalised result. A quantitative comparison will be made once the tensor-current renormalisation factor and the complete uncertainty budget are included. 

The $B_c^*-B_c$ hyperfine splitting is also of particular interest following the recent first observation of the ground-state $B_c^{*+}$ meson by the ATLAS Collaboration \cite{ATLAS:2026ubk}. ATLAS reports a mass splitting of $64.5\pm1.4\,(\mathrm{stat.})^{+1.0}_{-1.4}\,(\mathrm{syst.})$ MeV. From a ratio fit to the $B_c^*$ and $B_c$ correlators, we obtain
\begin{equation}
M_{B_c^*}-M_{B_c}=59.07(30)\ {\rm MeV},
\end{equation}
which is lower than the experimental value, but higher than earlier lattice determinations using NRQCD \cite{Gregory:2010gm, Dowdall:2012ab, Mathur:2018epb}, while being consistent with the more recent determination using an anisotropic clover action \cite{Cai:2026xja}. A more detailed study of the hyperfine splitting, including the full analysis of the ratio fits and associated systematic effects, will be presented in the forthcoming work.

Beyond the single-state analysis, the correlators contain additional information on excited-state contributions. As a complementary analysis, we employ the transfer-matrix generalised eigenvalue problem (TGEVP) approach of Ref.~\cite{Chakraborty:2024exj}. Starting from the same interpolating operator, the method constructs a Krylov basis and resolves several contributing states from the corresponding two-point correlator. We use the TGEVP results primarily as a cross-check of the ground-state determination and to examine excited-state and oscillating contributions. The states extracted with the TGEVP are then used to reconstruct the two-point correlators through the spectral form of Eq.~\eqref{eq:staggered_corr}, including both non-oscillating and oscillating contributions. 

\begin{figure}[h!]
    \centering
    \begin{subfigure}[t]{0.32\textwidth}
        \centering
        \includegraphics[width=\textwidth]{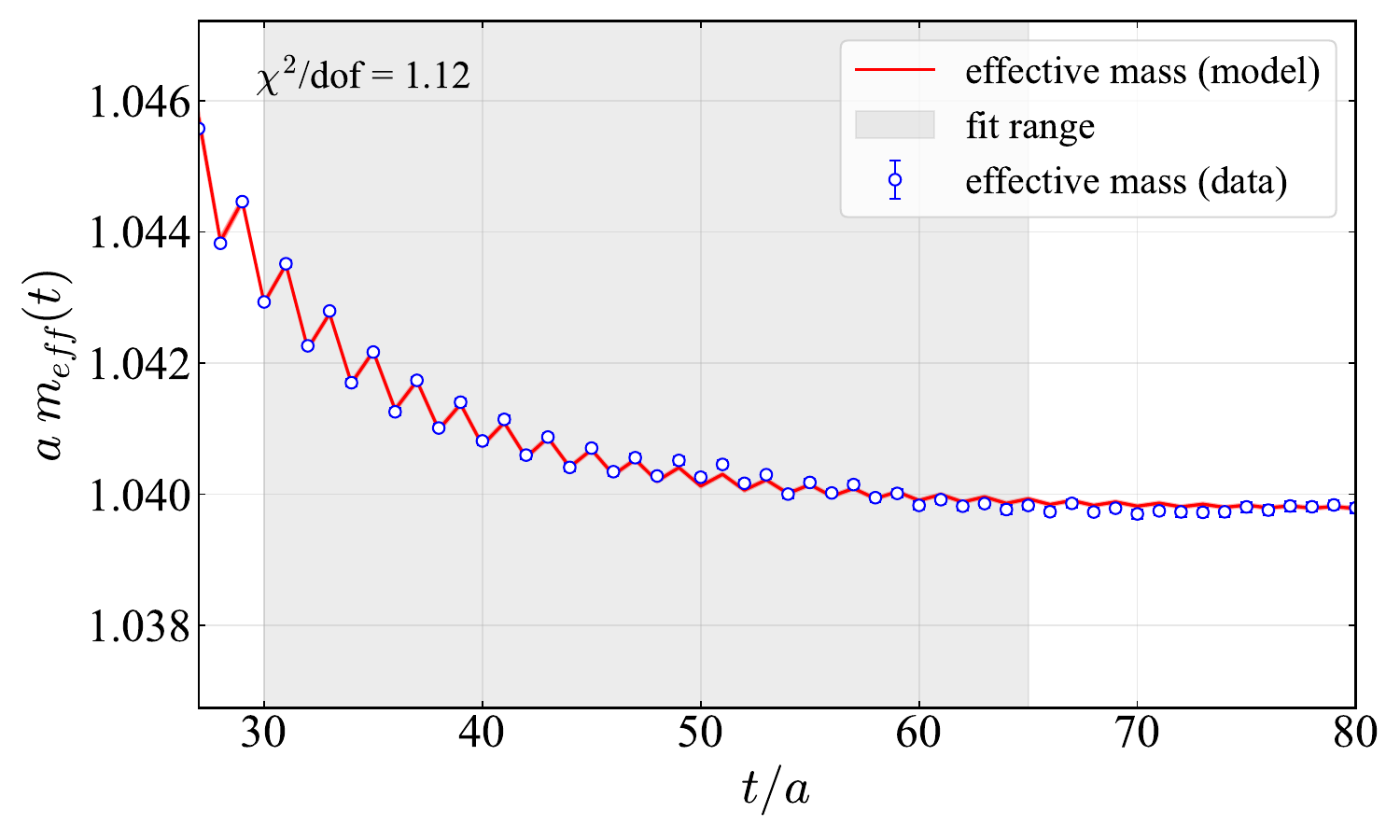}
        \caption{$B_c$ pseudoscalar channel.}
        \label{fig:tgevp_bc}
    \end{subfigure}
    \hfill
    \begin{subfigure}[t]{0.32\textwidth}
        \centering
        \includegraphics[width=\textwidth]{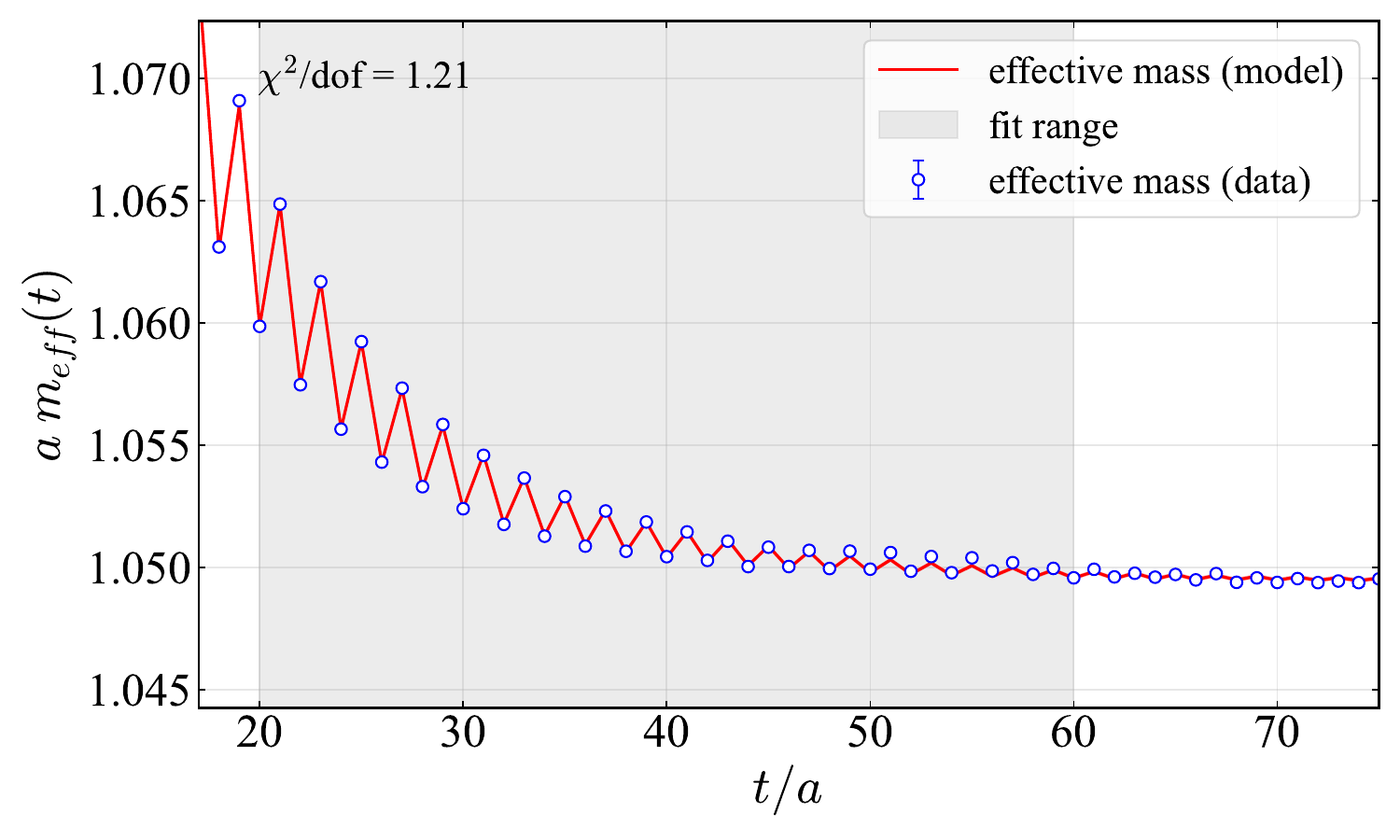}
        \caption{$B_c^*$ vector channel.}
        \label{fig:tgevp_bcstar}
    \end{subfigure}
   \hfill
    \begin{subfigure}[t]{0.32\textwidth}
        \centering
        \includegraphics[width=\textwidth]{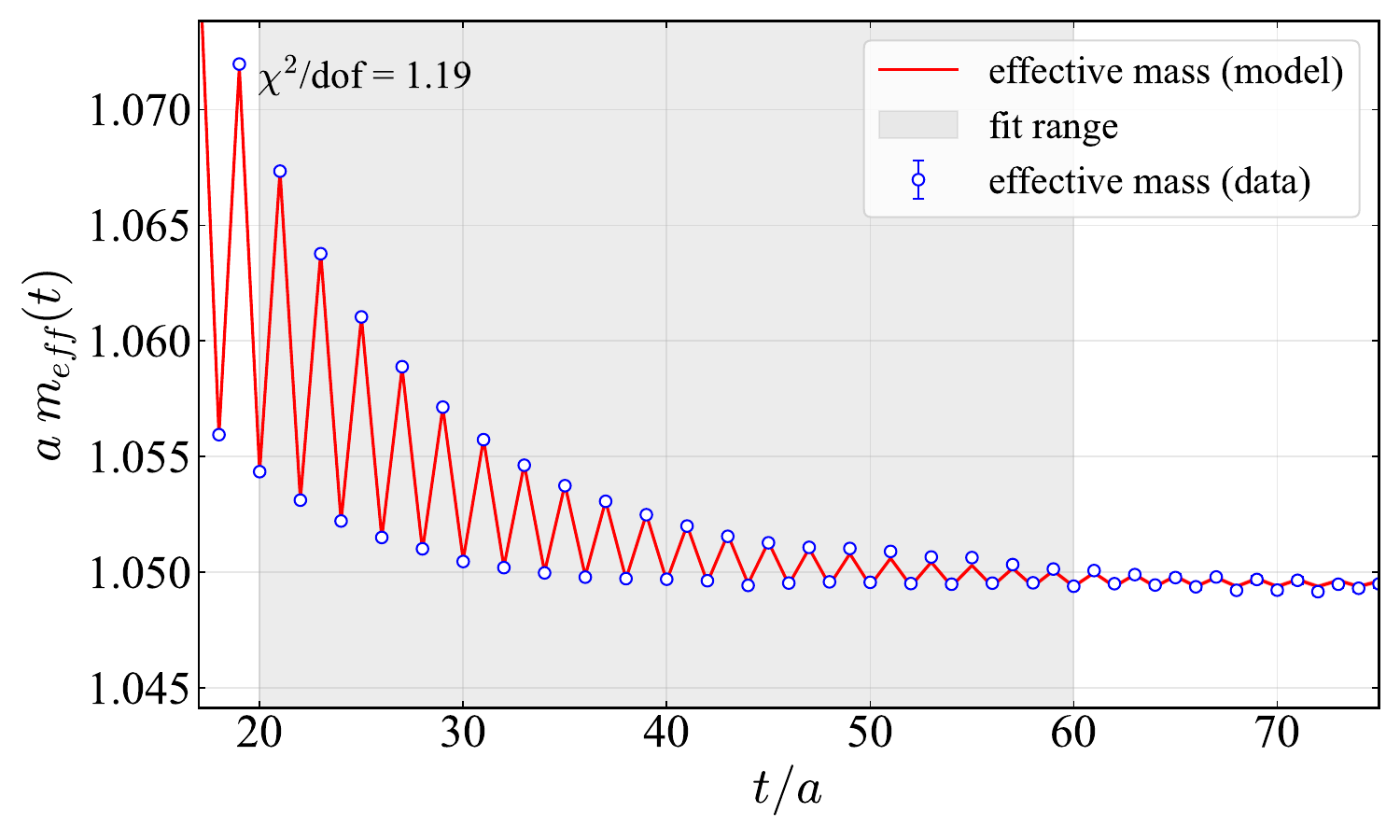}
        \caption{$B_c^*$ tensor channel.}
        \label{fig:tgevp_bcstar_tensor}
    \end{subfigure}

    \caption{ TGEVP analysis for the $B_c$, $B_c^*$ vector, and $B_c^*$ tensor channels, including both non-oscillating and oscillating state contributions.}

    \label{fig:tgevp}
\end{figure}

Results for the $B_c$ and $B_c^*$ channels using the TGEVP method are shown in Fig.~\ref{fig:tgevp}. The spectral ansatz includes four non-oscillating and three oscillating states. Ground-state information obtained from the late-time region is used to constrain the description at earlier Euclidean times, where excited-state contributions become more important. The reconstructed correlators reproduce the observed effective-mass behaviour over a broad time range, including the oscillatory structure at shorter separations. The ground-state and low-lying
excited-state quantities extracted from the TGEVP analysis are summarised in Table~\ref{tab:tgevp_results}.

\begin{table}[htbp]
\centering
\small
\renewcommand{\arraystretch}{1.55}
\begin{tabular}{lcccc}
\hline\hline
Channel
& Mass [MeV]
& Decay quantity
& Value [MeV]
& Excited states [MeV] \\
\hline

$\eta_c$
& $2984.10\left({}^{+0.19}_{-0.19}\right)$
& $f_{\eta_c}$
& $396.49\left({}^{+0.53}_{-0.63}\right)$
& $2S:\ 3675\left({}^{+60}_{-72}\right)$ \\
\hline

$\eta_b$
& $9406.78\left({}^{+0.11}_{-0.10}\right)$
& $f_{\eta_b}$
& $723.55\left({}^{+0.49}_{-0.51}\right)$
& $2S:\ 9993\left({}^{+53}_{-56}\right)$ \\
\hline

\multirow{2}{*}{$B_c$}
& \multirow{2}{*}{$6274.84\left({}^{+0.24}_{-0.26}\right)$}
& \multirow{2}{*}{$f_{B_c}$}
& \multirow{2}{*}{$451.24\left({}^{+0.77}_{-0.82}\right)$}
& $1P\,(0^+):\ 6709\left({}^{+13}_{-16}\right)$ \\
&
&
&
& $2S:\ 6893\left({}^{+83}_{-88}\right)$ \\
\hline

$B_c^*$ (vector)
& $6332.93\left({}^{+0.57}_{-0.59}\right)$
& $f_{B_c^*}/Z_V$
& $439.31\left({}^{+1.81}_{-1.87}\right)$
& $1P\,(1^+):\ 6746\left({}^{+22}_{-25}\right)$ \\
\hline

$B_c^*$ (tensor)
& $6333.08\left({}^{+0.57}_{-0.57}\right)$
& $f_{B_c^*}^{T}/Z_T$
& $396.27\left({}^{+1.63}_{-1.50}\right)$
& $1P\,(1^+):\ 6731\left({}^{+22}_{-30}\right)$ \\
\hline

\multirow{2}{*}{$\Upsilon$ (vector)}
& \multirow{2}{*}{$9460.79\left({}^{+0.13}_{-0.15}\right)$}
& \multirow{2}{*}{$f_{\Upsilon}/Z_V$}
& \multirow{2}{*}{$700.94\left({}^{+0.68}_{-0.72}\right)$}
& $1P\,(h_b):\ 9899.6\left({}^{+9.1}_{-8.5}\right)$ \\
&
&
&
& $2S:\ 10006\left({}^{+46}_{-44}\right)$ \\
\hline

\multirow{2}{*}{$\Upsilon$ (tensor)}
& \multirow{2}{*}{$9460.86\left({}^{+0.14}_{-0.14}\right)$}
& \multirow{2}{*}{$f_{\Upsilon}^{T}/Z_T$}
& \multirow{2}{*}{$646.81\left({}^{+0.61}_{-0.64}\right)$}
& $1P\,(\chi_{b1}):\ 9903.9\left({}^{+8.5}_{-11.3}\right)$ \\
&
&
&
& $2S:\ 9996\left({}^{+39}_{-44}\right)$ \\

\hline\hline
\end{tabular}

\caption{
Ground-state and low-lying excited-state quantities obtained from the TGEVP analysis.}
\label{tab:tgevp_results}
\end{table}

Some of the ground-state values obtained from the TGEVP analysis differ slightly from those extracted with the single-state fits. These differences are sensitive to the choice of fit ranges, in particular to the values of $t_{\min}$ adopted in the two analyses. A more detailed comparison of the fit-window dependence and the resulting central values will be included in the full analysis. The excited-state spectrum in these heavy-quark systems is already well established in lattice QCD. The single-operator TGEVP analysis used here provides a complementary cross-check of excited-state contamination and of the stability of the state extraction. We include these results to demonstrate the usefulness of the TGEVP approach as an independent analysis strategy and to demonstrate the robustness of the extracted ground- and excited-state energies.

\section{Summary}
\label{sec:summary}
We have presented a fully relativistic lattice-QCD study of the $B_c$ and $B_c^*$ systems using an ultrafine $N_f=2+1+1$ HISQ ensemble. The fine lattice spacing allows both charm and bottom quarks to be treated within the same relativistic formulation.

From the single-state analysis, we obtain the preliminary results
\begin{equation}
f_{B_c}=449.76(97)\ {\rm MeV},\qquad
\frac{f_{B_c^*}}{Z_V}=438.82(161)\ {\rm MeV},\qquad
\frac{f_{B_c^*}^{T}}{Z_T}=394.02(150)\ {\rm MeV},
\end{equation}
where the uncertainties shown are statistical only. The tensor channel constitutes the first lattice study of the $B_c^*$ tensor decay constant. The present analysis also gives a $B_c^*-B_c$ hyperfine splitting of $59.07(30)$ MeV from a ratio fit.

As a complementary analysis, we have applied the single-operator TGEVP method to investigate excited-state and oscillating contributions and to provide an independent cross-check of the ground-state extraction. The analysis also gives access to several low-lying excited states in the $B_c$, $B_c^*$, and bottomonium channels.

The vector and tensor quantities quoted here are prior to current renormalisation. The full analysis will incorporate additional statistics and the corresponding renormalisation factors, together with a detailed assessment of the systematic uncertainties associated with the correlator fits and the extraction of the decay constants and masses. These improvements will enable a more precise determination of the $B_c$ and $B_c^*$ decay constants and hyperfine splitting, while the TGEVP analysis will provide additional information on the low-lying excited-state spectrum.

\acknowledgments
This work is supported by the Department of Atomic Energy, Government of India, under Project Identification Numbers RTI-4012 and RTI-4013. Computations were carried out on the computing clusters at the Department of Theoretical Physics, TIFR, Mumbai. We are thankful to the MILC collaboration and, in particular, to Steven Gottlieb for providing us with the HISQ lattice ensembles. We would also like to thank Ajay Salve and Kapil Ghadiali for computational support.

\bibliographystyle{JHEP}
\bibliography{lattice26}
\end{document}